\documentstyle[11pt,paspconf]{article}

\begin{document}

\title{
 Polarisation Observations of Gravitational Lenses
  }
\author{
 A.~R. Patnaik$^1$, K.~M. Menten$^1$, R.~W. Porcas$^1$, A.~J. Kemball$^2$
  }
\affil{
$^1$Max-Planck-Institut f{\"u}r Radioastronomie, Bonn, Germany \\
$^2$National Radio Astronomy Observatory, Socorro, USA \\
  }

\begin{abstract}

  We present multi-frequency VLA polarisation observations of nine
  gravitational lenses. The aim of these observations was to determine
  Faraday rotation measures (RM) for the individual lensed images, and
  to measure their continuum spectra over a wide range of frequencies.

\end{abstract}


\section{Introduction}

Radio polarisation observations of gravitational lenses provide
important information about the properties of the lens as well as the
background radio source. Such observations give two measurable
quantities -- the degree and position angle (PA) of polarisation.  Both
are unaffected by the gravitational potential of the lens.  However,
the magneto-ionic medium in the lens can cause Faraday rotation of the
radiation, which may be different for each of the ray paths.
The magnitude of this differential rotation measure (RM) may provide
clues to the nature of the lensing galaxy; a gas-rich lens is expected
to give rise to a larger value than a gas-poor one. The polarisation
properties can also be used to discriminate between candidates in
surveys for gravitational lenses.  In addition, given that both the
degree and PA of polarisation in a compact radio source may vary in
time, permits independent measurements of the time delay.

There are, however, difficulties in the interpretation of polarisation
measurements, which arise mainly because many radio sources have
extended structure, and the spectral and polarisation characteristics change 
across the source. Polarisation variability, and the existence 
of time delays between images, may combine to make difficult a comparison of their
properties at a single observing epoch.

\section{Observations and Results}

We observed 9 radio lensed systems in which a compact core is multiply
imaged. The observations were made on 1998 May 22/23 using the NRAO
VLA in A-configuration in the 1.4, 5, 8.4, 15, 22 and 43-GHz bands.
Here we provide only a summary of our results; further details will be
published elsewhere.  References for individual sources are available
at the CASTLES web-site for gravitational lenses
(http://cfa-www.harvard.edu/glensdata/) maintained by C.S. Kochanek,
E.E. Falco, C. Impey, J. Leh{\'a}r, B. McLeod and H.-W. Rix.

{\bf B0218+357}~~Our fits of RMs to the present data yield
--8920$\pm$250~rad~m$^{-2}$ for image A, and
--7920$\pm$220~rad~m$^{-2}$ for B. These values differ from earlier
measurements (Patnaik et al., 1993).  However, a fit to the PA
difference between A and B at 15, 22 and 43 GHz gives a differential
RM of 980$\pm$10~rad~m$^{-2}$, similar to the value previously
reported.

{\bf MG0414+0534}~~This source is remarkably unpolarised, although
small but significant polarisation (0.2\%) is detected from the A1--A2
image complex at 5~GHz.  Image C was not detected at 43~GHz.

{\bf 0957+561}~~We find a RM for image A of
$-$61$\pm$1.0~rad~m$^{-2}$, and for B $-$91$\pm$1.0~rad~m$^{-2}$, with
equal intrinsic PAs (i.e. PA at zero wavelength).  For A our value
agrees with that given by Greenfield et al. (1985) but for B it differs
considerably from their $-$164.6$\pm$4.5~rad~m$^{-2}$.  This could
indicate a possible 180$^{\circ}$ PA ambiguity error at 1.4 GHz in the
earlier value, or a real change in RM along the path close to the
lensing galaxy G1.

{\bf B1422+231}~~The measured RMs of A, B and C are $-$4230$\pm$80,
$-$3440$\pm$80, $-$3340$\pm$80~rad~m$^{-2}$, respectively; their
intrinsic PAs are 90$^{\circ}\pm$10$^{\circ}$,
57$^{\circ}\pm$10$^{\circ}$ and 59$^{\circ}\pm$10$^{\circ}$.  It is
quite surprising that the RMs are so large, given that the lens galaxy
is reported to be an elliptical.

{\bf B1600+434}~~The RMs of the two images are low in this source, 44
and 40~rad~m$^{-2}$ for A and B, respectively. Comparing this source
to B0218+357, it is curious that the lens, a spiral galaxy, does not
give rise to large RM.

{\bf B1608+656}~~The images are unpolarised ($<$0.5\%) at all the
frequencies we observed.

{\bf PKS1830$-$211}~~This is the most difficult source to characterise,
as the PAs of the two images, measured at the brightness peaks at
each frequency, do not follow a $\lambda^2-$law. This result is not
surprising due to differing resolutions at different frequencies and
the frequency-dependent structures near the cores.

{\bf B1938+666}~~The three bright polarised emission regions, A, B and
C, have RMs of 665$\pm$90, 465$\pm$90 and
530$\pm$90~rad~m$^{-2}$, respectively. The source was not detected at
43~GHz.

{\bf 2016+112}~~The source was not detected in polarised emission. 

In summary, we detect polarised emission from 7 of the 9 lensed
systems.  The image flux ratios are generally independent of
frequency. We detect steepening of the spectra of many sources towards
high frequencies (e.g. 22 and 43~GHz).  Although the difference in RM
is expected to reflect the nature of the lensing galaxies, our results
do not clearly show this.  The lack of large RM in B1600+434 and
the presence of large RM in B1422+231 are especially puzzling.

\acknowledgments
The National Radio Astronomy Observatory is a facility
of the National Science Foundation operated under cooperative
agreement by Associated Universities, Inc.

\end{document}